\documentclass[11pt,twoside]{article}
\usepackage{./asp2014}
\usepackage{graphicx}
\newcommand*{\vcenteredhbox}[1]{\begingroup
\setbox0=\hbox{#1}\parbox{\wd0}{\box0}\endgroup}
\def\ltsim{\mathrel{\hbox{\rlap{\hbox{\lower4pt\hbox{$\sim$}}}\hbox{$<$}}}}

\aspSuppressVolSlug
\resetcounters

\begin{document}

\title{Magnetic fields of Be stars: preliminary results from a hybrid analysis of the MiMeS sample}
\author{G.A. Wade$^1$, V. Petit$^2$, J.H. Grunhut$^3$, C. Neiner$^4$ and the MiMeS Collaboration 
\affil{$^1$Dept. of Physics, Royal Military College of Canada, Kingston, ON, Canada K7K 7B4 \email{wade-g@rmc.ca}\\
$^2$Dept. of Physics \& Space Sciences, Florida Institute of Technology, Melbourne, FL, USA\\
$^3$ESO, Karl-Schwarzschild-Str. 2, D-85748 Garching, Germany\\
$^4$LESIA, Observatoire de Paris, CNRS, UPMC, Universit\'e Paris Diderot, Meudon, France
}
}

%J. Grunhut, V. Petit, C. Neiner, E. Alecian, Th. Rivinius, %S. Owocki, J. Landtreet, M. Shultz, R. Townsend, A. ud Doula
%& the MiMeS collaboration

% This section is for ADS Processing.  There must be one line per author.
\paperauthor{Gregg A. Wade}{wade-g@rmc.ca}{}{Royal Military College of Canada}{Department of Physics}{Kingston}{Ontario}{K7K 7B4}{Canada}

\begin{abstract}
In the context of the MiMeS survey of magnetism in massive stars, 85 classical Be stars were observed in circular polarization with the aim of detecting magnetic fields at their surfaces. No direct evidence of such fields is found, in contrast to the significant rate of detection (5-10\%) in non-Be B-type stars. In this paper we describe the sample properties, the methodology and the data quality. We describe a novel method, previously applied to Herbig Ae/Be stars, that allows us to infer upper limits on organized (dipolar) magnetic fields present in the photospheres of our targets. We review the characteristics and robustness of this null result, and discuss its implications.
\end{abstract}

\section{Introduction}

\begin{figure}
\centering
\vcenteredhbox{\includegraphics[width=2.5in]{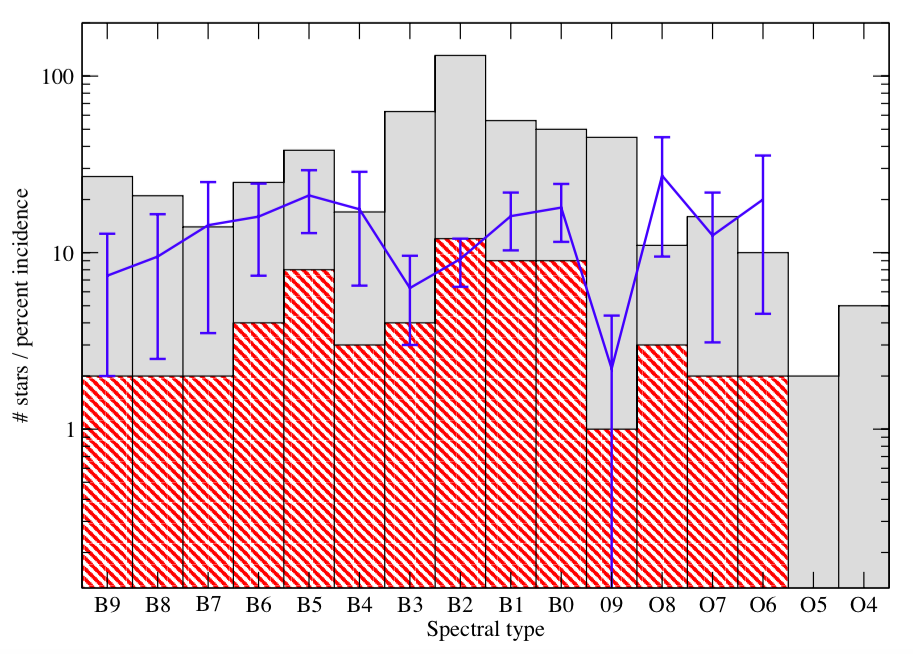}}\vcenteredhbox{\includegraphics[width=2.45in]{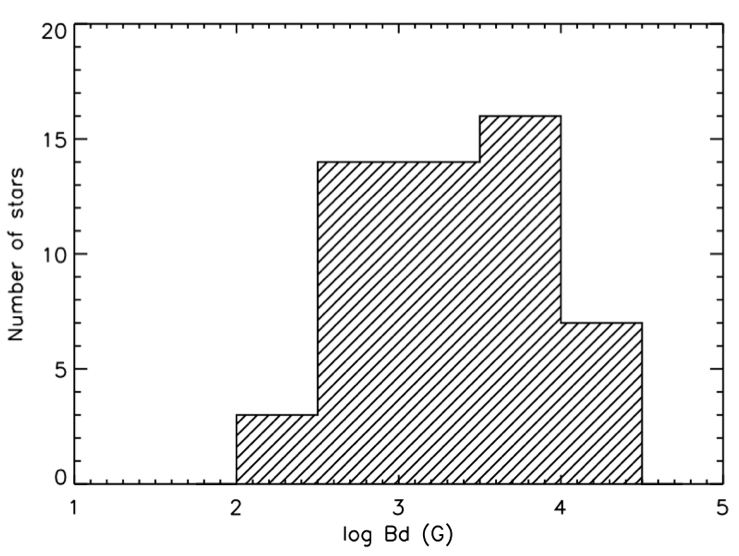}}
\caption{{\em Left -}\ The incidence of magnetic fields in B and O type stars as a function of spectral type \citep{2014IAUS..302..265W}. {\em Right -}\  Distribution of inferred surface dipole strengths of magnetic B-type stars from \citet{2013MNRAS.429..398P}.}
\label{magb}
\end{figure}

As a consequence of the impressive sensitivity of modern polarimetric instrumentation, coupled with dedicated large-scale surveys on intermediate- and large-aperture optical telescopes, a growing population of magnetic B-type stars is being identified. According to \citet{2013MNRAS.429..398P}, 54 magnetic B-type stars with effective temperatures above 16\,000~K are known. The magnetic fields of these stars have the following principal characteristics: they are strong (typically from a few hundred to a few thousand gauss at the stellar surface), organized (they have important dipole components that dominate the global field topology) and stable (they are not observed to change intrinsically on timescales of years to decades). They are observed to be present in a small fraction (between 5-10\%) of all B-type stars observed in the Magnetism in Massive Stars (MiMeS) survey \citep{2014IAUS..302..265W}. The incidence of magnetic stars versus spectral type, as well as the distribution of inferred magnetic dipole strengths, are illustrated in Fig.~\ref{magb}.

A defining physical feature of magnetic B stars is the interaction between the magnetic field and the outflowing radiatively-driven wind of the star. Although the winds of main sequence B-type stars are weak (i.e. characterized by mass loss rates typically below $10^{-8}~M_\odot$/yr), the ability of the magnetic field to channel, retain and concentrate wind plasma is sufficiently strong in most magnetic B stars to modify the profiles and variability of wind-sensitive ultraviolet resonance lines in important and characteristic ways \citep[e.g.][]{huib2012}. Under the additional influence of supporting centrifugal forces provided by rapid rotation, the concentration of plasma is significantly enhanced. In the case of such a "centrifugal magnetosphere" (CM), optical diagnostics - in particular H$\alpha$ - show rotationally modulated emission that is characteristic of the overall structure and density distribution of the magnetosphere. Models \citep[e.g.][]{townsend05} and observations \citep{townsendsigori,carciofi13,tomography} indicate that the gas distribution of such a magnetosphere is highly flattened, consisting of a thin disc-like distribution of plasma in the magnetic equatorial plane, in addition to "clouds" or "spokes" of enhanced density located at the intersections of the magnetic and rotational equatorial planes. 

As discussed by \citet[][]{2012MNRAS.426.2738N,2014arXiv1411.2542S} and Owocki, these proceedings, such CMs are distinguished from the discs of classical Be stars in a number of important ways: Be star discs are known to be in Keplerian orbits, whereas CMs are in enforced corotation with the star; Be star discs are located approximately in the rotational equatorial plane, whereas CMs are located in an inclined magnetic equatorial plane; Be star discs undergo significant formative and dissipative episodes, whereas no such variability has been observed in a CM \citep[e.g.][]{town2013}.

Notwithstanding these differences, it seems natural to wonder whether the discs of classical Be stars could be launched, supported or structured by magnetic fields similar to those described above. Although from theoretical considerations it seems unlikely that large-scale fields could be responsible for raising material from the surface of a Be star into a Keplerian orbit \citep{2006ASPC..355..219O}, indirect observational evidence exists that is suggestive of the presence of magnetic fields at the surfaces or in the immediate environments of some Be stars. Candidates of particular interest are described in the next Section.

In this paper we describe a novel analysis aimed at testing the idea that organized magnetic fields are associated with Be disc formation, and implicitly examining theoretical predictions of the exclusivity of magnetic fields and the Be phenomenon \citep{2006ASPC..355..219O}. Our analysis is performed using spectropolarimetric observations of 85 classical Be stars obtained in the context of the MiMeS survey. 

\section{Magnetic Be star candidates in the literature}

Several notable historical reports of magnetic fields in classical Be stars exist.

The B3IIIe star $\omega$~Ori has long been reported to exhibit multiperiodic variability that has been ascribed to both nonradial pulsation and rotational modulation \citep{2001MNRAS.327.1288B, 2002A&A...388..899N}. \citet{2003A&A...409..275N} reported the marginal detection of a magnetic field in $\omega$~Ori. They interpreted their observation of the longitudinal magnetic field of this star in the context of the magnetic oblique rotator model, and inferred a surface dipole strength of $530\pm 230$~G and a rotational period of 1.29 days. However, an attempt to directly confirm and better characterize the magnetic field yielded null results, and an upper limit of 80~G on the strength of any dipole field that is present \citep{2012MNRAS.426.2738N}. However, indirect indications of the presence of a field remain.

$\beta$~Cep was reported by \citet{2001MNRAS.326.1265D} and \citet{2013A&A...555A..46H} to host a weak ($\sim 300$~G) dipole magnetic field. The field detection was clearly confirmed based on high-quality observations of Stokes $V$ Zeeman signatures measured over multiple cycles of the star's 12d rotational period. However, \citet{2006A&A...459L..21S} demonstrated that the H$\alpha$ emission from this binary system is in fact associated with the (apparently non-magnetic) secondary star.

$\gamma$~Cas and a small number of other stars of this class have also been claimed to host magnetic fields linking the star and the Be disc \citep[e.g.][]{2006ApJ...647.1375S}. However, to date no direct indications of such fields have been obtained.

A number of other Be stars that have been claimed to be magnetic in the literature have been found to be non-magnetic on further examination. Most of these claims were based on low-significance detections obtained from Stokes $V$ spectra acquired with the FORS1 and FORS2 instruments at VLT. The majority are discussed by \citet{2012A&A...538A.129B}, who attribute these false detections to important contributions of noise from non-photon sources such as small instrument flexures. 

\begin{figure}
\centering
\includegraphics[width=4.3in]{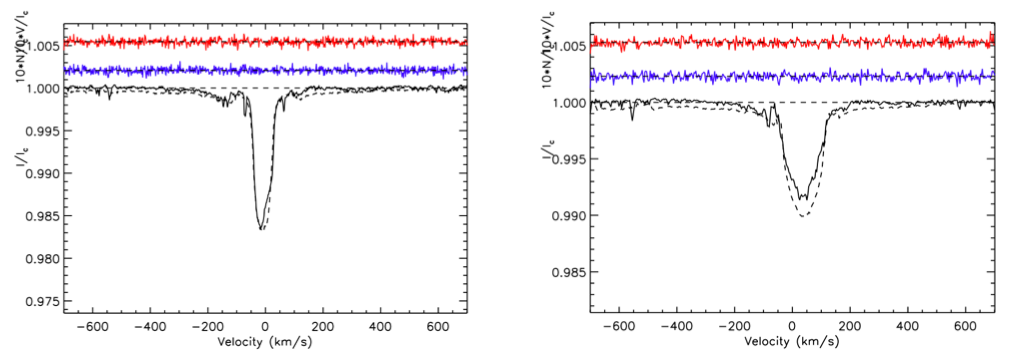}
\includegraphics[width=4.3in]{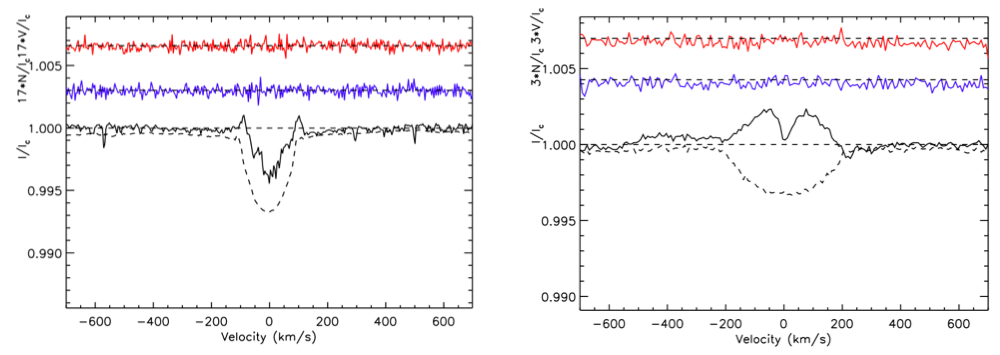}
\caption{Illustrations of observed versus synthetic LSD Stokes $I$ (black) profiles, as well as Stokes observed $V$ (red) and diagnostic null (blue) profiles for stars with profiles dominated by the photospheric component (upper frames) and with significant CS contamination (lower row). For the Stokes $I$ profiles, the full line is the photospheric profile extracted from the true observed spectrum, while the dashed line is the profile extracted from the synthetic spectrum. }
\label{lsd}
\end{figure}

\section{The MiMeS survey of classical Be stars}

The MiMeS survey examined high resolution circular polarization observations of approximately 550 stars obtained with the ESPaDOnS, Narval and Harpspol spectropolarimeters. Of these, about 450 were B-type stars and $\sim 100$ were initially identified as emission-line stars. After a careful examination of the emission-line sample, 213 spectra of 85 classical Be stars were identified. Classification as a classical Be stars was verified using the BeSS database, and retaining only those stars with a positive classification. 

%As a consequence, 15-20 stars were rejected from the sample before analysis.

The data quality is very good, with a median SNR of 800 per 1.8 km/s pixel.

\section{Procedure}

Stellar magnetic fields are most often detected and characterized using measurements of small circular polarization (Stokes $V$) variations across photospheric spectral lines. In the presence of an external magnetic field, atomic electronic transitions become split in a manner that is sensitive to both the field strength and its local orientation. Circular polarization, which is produced as a consequence of a line-of-sight, or longitudinal, component of the magnetic field, represents the most sensitive field diagnostic. The presence of a magnetic field in the stellar photosphere produces characteristic signatures ("Zeeman signatures") with similar shapes, but different amplitudes, in many spectral lines. Modern multi-line methods such as the widely-used Least-Squares Deconvolution \citep[LSD; ][]{1997MNRAS.291..658D,2010A&A...524A...5K} exploit this self-similarity to co-add profiles, improving the signal-to-noise ratio significantly. 

%\begin{figure}
%\centering
%\vcenteredhbox{\includegraphics[width=3in]{mag_det1.png}}\vcenteredhbox{\includegraphics[width=2in]{mag_det2.png}}
%\caption{Measurement of stellar magnetic fields using high resolution circular polarization spectroscopy. {\em Left -}\ A small region of a circularly polarized spectrum of the magnetic B0.5V star HD~63425. The lower (black) curve shows the unpolarized (Stokes $I$) spectrum, while the upper (red) curve shows the (magnified and shifted) circular polarization (Stokes $V$) spectrum. Note the weak polarization variation across each spectral line - the signature of the presence of a photospheric magnetic field. {\em Right -}\ The mean Stokes $I$ and $V$ line profiles of HD~63425 computed using the multiline method Least-Squares Deconvolution \citep{1997MNRAS.291..658D,2010A&A...524A...5K}. Note the improvement in the significance of the magnetic detection, which in this case corresponds to a longitudinal magnetic field of $130\pm 10$~G.}
%\label{magnetic}
%\end{figure} 

Be stars suffer from contamination of their photospheric spectra due to emission from the circumstellar (CS) environment. In the hypothetical case considered here, in which the star is strongly magnetized but the disc magnetic field is negligible (due to the rapid $r^{-3}$ decay of the large-scale stellar field), this contamination modifies the intensity (Stokes $I$) spectrum, but leaves the Stokes $V$ spectrum essentially unchanged. 

CS emission impacts magnetometry in two ways. First, the modification of the Stokes $I$ profile can make it difficult to assess the extent of the photospheric Stokes $I$ profile. This makes it more challenging to evaluate the potential presence of weak signal in Stokes $V$. Moreover, modification of the Stokes $I$ profile shape and equivalent width make it impossible to accurately measure the longitudinal magnetic field. In the absence of a magnetic detection, the longitudinal field error bar is essentially the only important field-related statistical quantity, as it provides an estimate of the upper limit on admissible fields. Because this error bar is sensitive to CS contamination of Stokes $I$, it is important to understand, and potentially limit, the CS contribution to the $I$ profile (even if the $V$ profile is unmodified).

As a consequence of these effects, magnetometry of Be stars is inherently less accurate that for stars with predominantly photospheric spectra. As a solution to this problem, we adopt the 'hybrid' approach of \citet{2013MNRAS.429.1001A}, replacing the real (CS-contaminated) LSD Stokes $I$ profile of each Be star with one computed from a synthetic spectrum. We used the Synth3 LTE spectrum synthesis code \citep{2007pms..conf..109K}, and effective temperatures from the published literature or estimated from the BeSS database. The $v\sin i$ and $v_{\rm rad}$ of each star were obtained from examination of the cleanest spectral lines (typically high excitation lines) in their spectra. We computed each star's photospheric Stokes $I$ spectrum with the same spectral domain and resolution as the observed spectra, assuming solar abundances and adding synthetic gaussian noise. For each observed spectrum we then used the full LSD line mask appropriate to the star to extract the Stokes $I$ LSD profile from the synthetic spectrum, and the Stokes $V$ LSD profiles from the observed spectrum. Combining the $I$ and $V$ profiles, this ultimately resulted in hybrid LSD profiles. The advantage of this approach is that we avoid the uncertainty related to CS contributions to the Stokes $I$ profile. 

Illustrations of the raw and hybrid LSD profiles of several stars in our sample are shown in Fig.~\ref{lsd}. This figure shows the good agreement between the observed and computed Stokes $I$ profiles when CS contamination is low.

From the hybrid profiles we extract two measurements: the likelihood of detection of a significant signal in the Stokes $V$ profile across the inferred photospheric spectral line (i.e. the detection probability), and the longitudinal magnetic field.

\section{Results}

\begin{figure}
\centering
\vcenteredhbox{\includegraphics[width=2.35in]{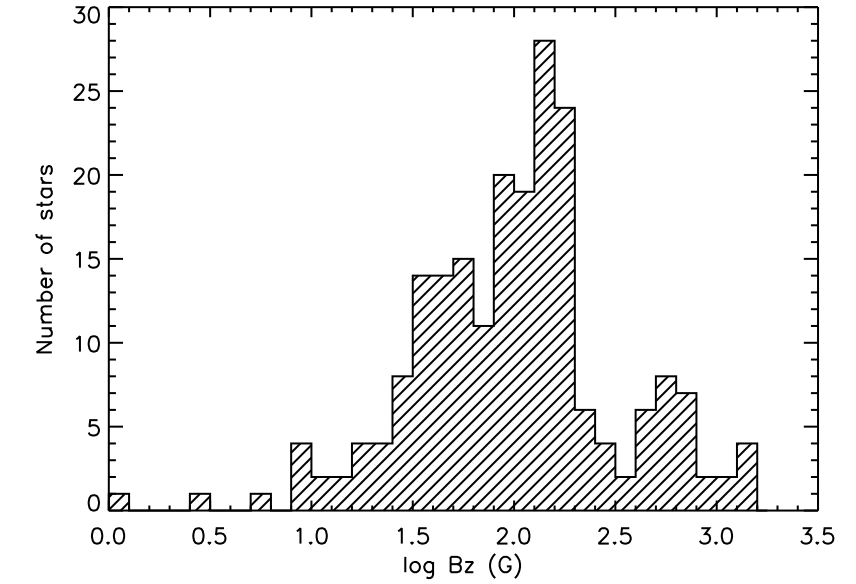}}\vspace{0.5cm}\vcenteredhbox{\includegraphics[width=2.15in]{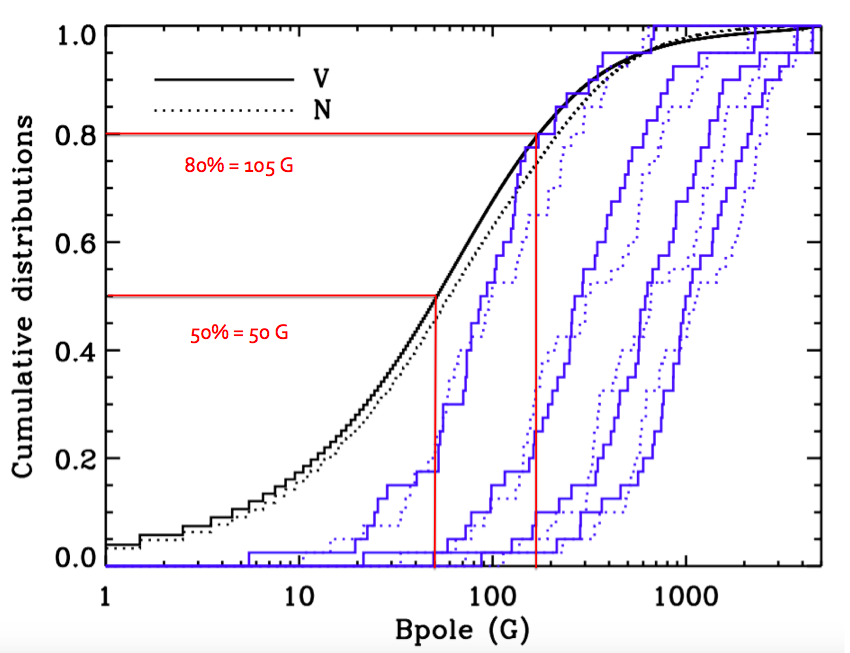}}
\caption{{\em Left -}\ Distribution of longitudinal magnetic field $1\sigma$ uncertainties for the $\sim 85$ Be stars analysed using the hybrid approach. {\em Right -}\ Inferred distribution of surface dipole field upper limits for the $\sim 50$\% of the sample analysed using the method of \citet{2012MNRAS.420..773P}. The leftmost curve shows the field strength distribution built directly from the probability density functions of individual Be stars. The derived median and 80th percentile most probable field strengths are 50~G and 105~G, respectively. The 4 curves on the right side show the cumulative distributions built from the upper limits of the 68.3, 95.4, 99.0, and 99.7 percent credible regions, for both Stokes $V$ and the diagnostic null. }
\label{sigmas}
\end{figure}

No evidence of magnetic field is detected in any of the studied stars, either according to the detection probability or the longitudinal magnetic field. Fig.~\ref{sigmas} (left panel) illustrates the distribution of derived longitudinal magnetic field formal (i.e. $1\sigma$) uncertainties. The median uncertainty is 103~G. In our sample, 27 spectra (corresponding to 17 stars) yield uncertainties $\leq 30$~G, and 8 spectra (corresponding to 7 stars) yield uncertainties $\leq 10$~G.

We have applied the method of \citet{2012MNRAS.420..773P} to about 1/2 of the sample and calculated the sample's dipole field strength probability distribution function, as well as the cumulative distributions of the field upper limits. For the sample studied, the results imply that the dipole field components of 50\% of the stars are probably weaker than 50~G, whereas those of 80\% of the sample are probably weaker than 105~G. These limits should be compared to the distribution of dipole field strengths of known magnetic B stars in Fig.~\ref{magb}.

Two of the rapidly rotating main sequence B stars showing H$\alpha$ emission that were removed from our sample before the analysis are HR 5907 and HR 7355. Both of these stars exhibit strong magnetic fields \citep{2010MNRAS.405L..46R,2010MNRAS.405L..51O,2012MNRAS.419.1610G} with respective surface dipole strengths of 10-16 kG and 13-17 kG. Obviously, magnetic fields of this magnitude would have been easily detected in most stars of the sample, both according to the longitudinal field and the analysis following \citet{2012MNRAS.420..773P}. 

However, HR 5907 and HR 7355 were removed from the sample of classical Be stars. In the case of HR 7355, the star does not appear in the BeSS database. HR 5907 is indeed listed in BeSS, and was considered to be a classical Be star until the discovery of its magnetic field. However, examination of the H$\alpha$ profiles of these stars shows that they are clearly incompatible with emission from a disc in Keplerian rotation, as per the definition of \citet{2013A&ARv..21...69R}. In fact, these two stars have been demonstrated to exhibit centrifugal magnetospheres, similar to other known rapidly rotating magnetic B stars that exhibit emission yet are not considered to be Be stars \citep[e.g.][]{2012MNRAS.419..959O,2013MNRAS.429..398P}.

%HR7355 is not in BeSS and is not claimed as a classical Be star. HR5907 however is indeed listed in BeSS, as it was considered a classical Be star before the field was discovered.

 \section{Summary and conclusions}
 
\begin{figure}
\centering
\vcenteredhbox{\includegraphics[width=3.2in]{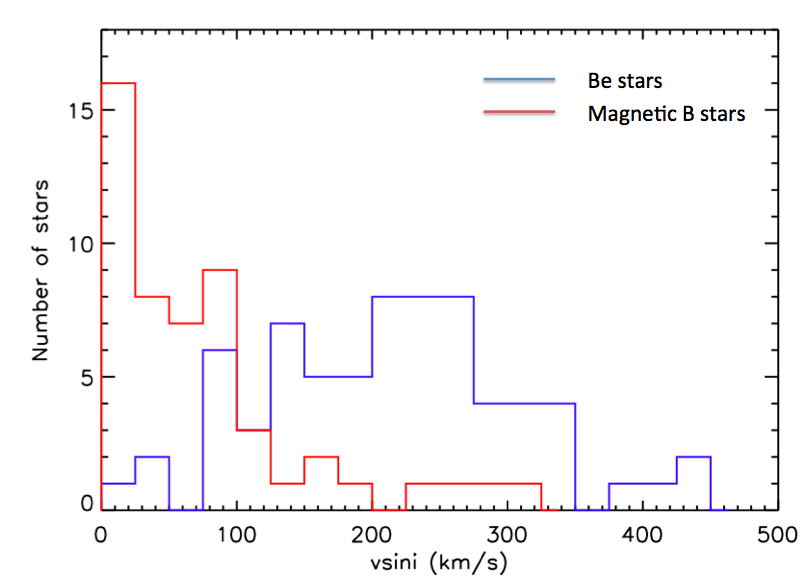}}
\caption{The projected rotational velocities of the MiMeS Be stars sample compared to those of the known magnetic B stars.}
\label{bevsbp}
\end{figure}

Our hybrid analysis of the MiMeS Be sample yields no detection of magnetic fields. We conclude that classical Be stars and B stars with strong, organised magnetic fields appear to be mutually exclusive populations. This is supported by their very different distributions of rotational velocities, as illustrated in Fig.~\ref{bevsbp} for the MiMeS Be stars sample and the sample of known magnetic B stars summarized by \citet{2013MNRAS.429..398P}. In particular, the MiMeS Be star sample shows a broad $v\sin i$ distribution with a peak near 250~km/s, and extending to over 400~km/s. The magnetic B stars show a distribution that peaks at low $v\sin i$ ($\ltsim 50$~km/s), with a small fraction of the sample having $v\sin i>100$~km/s.

Intuitively, it would seem that magnetic fields provide a natural mechanism for removing magnetic stars from the population of Be stars. Magnetic braking slows stellar rotation, and thereby will naturally remove magnetic stars from the rapidly-rotating population that constitute the Be stars on the spindown timescale \citep{2009MNRAS.392.1022U,2010ApJ...714L.318T}. Perhaps an examination of the youngest populations of B stars could provide a fruitful way to find magnetic Be stars, since such objects would not have suffered much magnetic braking.

Nevertheless, Fig.~\ref{bevsbp} shows that rapidly rotating magnetic B stars do exist. Perhaps such objects are examples of stars that might have been Be stars in the absence of their magnetism. However, even if this hypothesis is correct, it is still unclear by what mechanism(s) the Be phenomenon is interrupted. For example, the field may have a direct impact on disc formation through magnetically-enforced corotation. On the other hand, the field might indirectly affect other processes responsible for disc launching, such as pulsation.

%The current picture of the Be phenomenon today is that you need rapid rotation and pulsations to create a Be star. Bn stars are rapidly rotating but without pulsations and thus not Be. Whether there is a field or not probably doesn't make a difference. What is important is whether you can transport angular momentum from the core to the surface.

The modelling described in this paper has sought evidence of the presence of organized magnetic fields with important dipole components. Kochukhov \& Sudnik (2013) evaluated the potential detectability of complex magnetic fields, characterized by random bipolar spot distributions, in the photospheres of hot stars. Such an investigation applied specifically to Be stars might be of future interest.

\bibliography{wade}  % For BibTex

\section{Questions}

{\bf Atsuo Okazaki:} If we assume that the magnetic pressure is saturated at, say, 10\% of the gas pressure, the magnetic field is strongest in the innermost region of the Be disk and decreases rapidly outward, because of the rapid decrease of the disk gas pressure. Is there a way to detect such magnetic fields?

\noindent {\bf Gregg Wade:} In principle yes, but to evaluate detectability would require a prediction of the strength and topology of the predicted field (essentially its power spectrum).

\noindent {\bf Stan Owocki:} The lack of large scale fields in Be stars is consistent with
        the notion that such fields are inherently incompatible with
        the shearing in a Keplerian disk.  
        
\noindent {\bf Gregg Wade:} Agreed, although \citet{2012MNRAS.426.2738N} propose a scenario for $\omega$~Ori in which the physical scales of the Be disc and a putative magnetosphere are very different, in particular with the magnetosphere situated between the star and the inner disc radius.

\end{document}